\documentclass[pre,aps,showpacs,superscriptaddress,twocolumn]{revtex4-1}

\usepackage{amsmath}
\usepackage{amssymb}
\usepackage{amsbsy}
\usepackage{graphicx}
\usepackage{color}
 
\begin{document}
 
\title{Sample-to-sample fluctuations of power spectrum of a random motion in a periodic Sinai model}

\author{David S. Dean}
\affiliation{Universit\'e Bordeaux and CNRS,
  Laboratoire Ondes et Mati{\`e}re d'Aquitaine (LOMA), UMR 5798,
  F-33400 Talence, France}
\author{Antonio Iorio}
\affiliation{Dipartimento di Fisica, Sapienza Universit{\`a} di Roma,
  P.le A. Moro 2, I-00185 Roma, Italy}
\affiliation{Dipartimento di Matematica e Fisica, Universit{\`a} Roma Tre,
   Via della Vasca Navale 84, I-00146 Roma, Italy}
\author{Enzo Marinari}
\affiliation{Dipartimento di Fisica,
  Sapienza Universit{\`a} di Roma, P.le A. Moro 2, I-00185 Roma,
  Italy}
\affiliation{INFN, Sezione di Roma 1 and Nanotech-CNR, UOS di
  Roma, P.le A. Moro 2, I-00185 Roma, Italy}
\author{Gleb Oshanin}
\affiliation{Sorbonne Universit\'es, UPMC Univ Paris 06, UMR 7600,
  LPTMC, F-75005, Paris, France}
\affiliation{CNRS, UMR 7600,
  Laboratoire de Physique Th\'{e}orique de la Mati\`{e}re
  Condens\'{e}e, F-75005, Paris, France}

\date{\today}

\begin{abstract}
  The Sinai model of a tracer diffusing in a quenched Brownian
  potential is a much studied problem exhibiting a logarithmically
  slow anomalous diffusion due to the growth of energy barriers with
  the system size. However, if the potential is random but periodic,
  the regime of anomalous diffusion crosses over to one of normal
  diffusion once a tracer has diffused over a few periods of the
  system. Here we consider a system in which the potential is given by
  a Brownian Bridge on a finite interval $(0,L)$ and then periodically
  repeated over the whole real line, and study the power spectrum
  $S(f)$ of the diffusive process $x(t)$ in such a potential.  We show
  that for most of realizations of $x(t)$ in a given realization of
  the potential, the low-frequency behavior is
  $S(f) \sim {\cal A}/f^2$, i.e., the same as for standard Brownian
  motion, and the amplitude ${\cal A}$ is a disorder-dependent random
  variable with a finite support.  Focusing on the statistical
  properties of this random variable, we determine the moments of
  ${\cal A}$ of arbitrary, negative or positive order $k$, and
  demonstrate that they exhibit a multi-fractal dependence on $k$, and
  a rather unusual dependence on the temperature and on the
  periodicity $L$, which are supported by atypical realizations of the
  periodic disorder.  We finally show that the distribution of
  ${\cal A}$ has a log-normal left tail, and exhibits an essential
  singularity close to the right edge of the support, which is related
  to the Lifshitz singularity.  Our findings are based both on
  analytic results and on extensive numerical simulations of the process $x(t)$.
\end{abstract}


\maketitle

The statistical classification of time dependent stochastic processes
is often based on the study of their power spectrum
\begin{align}
\label{spec1}
S(f) = \lim_{T \to \infty} \overline{\left| \int^T_0 dt\; \,
  e^{i f t}\; x(t)\right|^2}\,,
\end{align} 
where the horizontal bar denotes ensemble averaging with respect to
all possible realizations of $x(t)$.  Many processes, which are
common in nature and are often observed in engineering and
technological sciences, are found to exhibit a low-frequency noise spectrum of
the universal form \cite{man,dutta}
 \begin{align}
 \label{spec2}
 S(f) \sim \dfrac{{\cal A}}{f^{\alpha}} \,.
 \end{align}
 The amplitude ${\cal A}$ is independent of $f$, and the exponent
 $\alpha \in (1, 2)$, with the extreme cases $\alpha = 1$ and
 $\alpha = 2$ corresponding to the $1/f$ (flicker) noise and Brownian
 noise (or noise of the extremes of Brownian noise \cite{gleb}),
 respectively.  There exist a few physical cases for which the form in
 \eqref{spec2} with $\alpha < 2$ extends over many decades in
 frequency, implying the existence of correlations over surprisingly
 long times. Relevant examples include electrical signals in vacuum
 tubes, semiconductor devices and metal films \cite{man,dutta}. More
 generally, the form in \eqref{spec2} is observed in sequences of
 earthquakes \cite{5} and weather data \cite{wea}, in evolution
 \cite{7}, human cognition \cite{8}, network traffic \cite{9} and even
 in the temporal distribution of loudness in musical recordings
 \cite{press}.  Recent experiments have shown the occurrence of such
 universal spectra in processes taking place in a variety of nanoscale
 systems. Among them are transport in individual ionic channels
 \cite{10,11} and electrochemical signals in nanoscale electrodes
 \cite{13}, bio-recognition processes \cite{14} and intermittent
 quantum dots \cite{15}.  Many other examples, related theoretical
 concepts, emerging challenges and unresolved problems have been
 discussed in \cite{15,enzo,enzo1,mike,16,17}.
 
An example of a transport process which exhibits the flicker $1/f$ noise (with
logarithmic corrections) was pointed out more than
30 years ago in \cite{enzo,enzo1}.  This is a paradigmatic example for
random motion in a quenched random environment, now known as Sinai
diffusion \cite{sinai}, which has been studied in many different
contexts \cite{bou1,osh1,osh2,mon,alb,com,sid}. Sinai diffusion is
defined as a Brownian motion advected by a quenched drift which is
time independent and uncorrelated in space. It can thus be seen as an
over-damped Langevin process subject to a quenched force which is uncorrelated in space, so that  in one dimension it is derived from a Brownian
potential $V(x)$. The mean-square displacement of the Sinai diffusion
exhibits a remarkably slow logarithmically growth with time $t$,
\begin{align}
\label{sinai}
\mathbb{E}\left(\overline{x^2(t)}\right)\sim \ln^4(t)\;, \,\,\, t \to \infty \,,
\end{align}
where $\mathbb{E}(\cdot)$ denotes averaging over realizations of the
random potential. The result in \eqref{sinai} is
supported by \textit{typical} realizations of disorder,{\em i.e.}, it
holds for almost all samples with a given potential $V(x)$.  Note that
despite the slow logarithmic dispersion of the trajectories, the
probability currents $J_L$ through finite samples of Sinai chains of
length $L$ appear to be much larger than the Fickian currents in
homogeneous systems \cite{osh1,osh2,mon,alb}; for finite Sinai chains
one has $\mathbb{E}(J_L) \sim 1/\sqrt{L}$, while for homogeneous
systems  $J_L \sim 1/L$. Such an anomalous behavior
of currents is supported by rare \textit{atypical} realizations of
$V(x)$ which however produce the dominant contributions to the
average.

\begin{figure}[ht]
\begin{center}
\centerline{\includegraphics[width = .45 \textwidth]{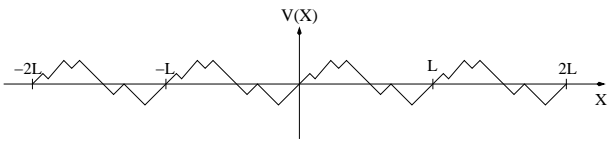}}
\caption{Potential $V(x)$ as a periodically extended Brownian Bridge
  with $V(x=0)=V(x=L)=0$.
\label{Fig1}
}
\end{center}
\end{figure} 

In this paper we analyze the power spectrum of random motion in a
random quenched potential looking at the problem from a different
perspective - we will mainly focus on the amplitude ${\cal A}$ of the
power spectrum, not on the value of the exponent $\alpha$
characterizing the power spectrum.  In random environments, this
amplitude is itself a random variable fluctuating from realization to
realization of the random potential, this makes the power spectrum
itself a random variable. Here we concentrate on a particular model -
a periodic Sinai chain \cite{dean}, in which the potential is a finite
Brownian trajectory with constrained endpoint - the so-called
 Brownian Bridge, defined on the interval $(0,L)$ and then
periodically extended  in both directions to give an infinite
one-dimensional system (see Fig.\ref{Fig1}). The origin of the slow logarithmic growth in
the original Sinai model (with $L = \infty$) is due to the unlimited growth of the Brownian potential and the associated energy barriers, however  in our periodic case $x(t)$
ultimately converges to a Brownian motion, on large time and length scales, so that the low frequency spectrum has a form in \eqref{spec2} with $\alpha = 2$ but the
amplitude ${\cal A}$ - a positive random variable with a finite
support $(0,{\cal A}_r)$ - fluctuates from sample to sample.  We
determine the moments of ${\cal A}$ and show that the probability
distribution function $P({\cal A})$ has a rather non-trivial form
characterized by a log-normal left tail (in the vicinity of $0$) and a
singular right tail (in the vicinity of the right edge ${\cal A}_r$ of
the support).  In general, ${\cal A}$ is not self-averaging and its
moments are supported by atypical realizations of disorder. These
analytic predictions for the periodic Sinai model are confirmed by
extensive numerical simulations. An analysis of the distribution of
${\cal A}$ for the original Sinai model with $L \equiv \infty$, where
the spectrum is described by \eqref{spec2} with $\alpha = 1$
\cite{enzo,enzo1} will be presented elsewhere.

The precise definition of the model studied is as follows.  Consider
the Langevin dynamics of a tracer $x(t)$ in a time-independent
potential $V\left(x\right)$:
\begin{equation}
\label{Langevin}
\eta \dfrac{d x(t)}{d t} = - \dfrac{d V\left(x(t)\right)}{d x(t)} + \xi_t \,,
\end{equation}
where $\eta$ is the friction coefficient, $\xi_t$ is a Gaussian white
noise with zero mean and covariance
\begin{equation}
\overline{\xi_t \xi_{t'}} = 2 \eta T \delta(t - t') \,,
\end{equation}
and $T$ is the temperature in units of the Boltzmann constant. The
potential is periodic, such that $V\left(x + L\right)=
V\left(x\right)$, with $L$ being the periodicity.

Furthermore, we assume that the potential $V(x)$ on the interval $x
\in (0,L)$ is a stochastic, continuous Gaussian process, pinned at
both ends so that $V(0) = V(L) = 0$, having zero mean and covariance
\begin{equation}
\mathbb{E}\left(V(x) V(y) \right) = {2 D_V}\left[{\rm min}(x,y) - 
\frac{xy}{L}\right] \,, \,\, 0 \leq x,\,  y \leq L \,,
\end{equation}
where $D_V = V_0^2/(2 l)$, $V_0$ being a characteristic extent of the
potential on a small scale of size $l$. In other words, $V(x)$ on the
interval $(0,L)$ is the so-called Brownian Bridge  (BB in what follows) \cite{mor} which has
the representation
\begin{equation}
V(x) = W_x - \frac{x}{L} W_L,
\end{equation}
where $W_x$ is a standard Brownian motion started at $W_0=0$ with
correlation function
\begin{equation}
\mathbb{E}\left(W_x W_y\right) = 2D_V {\rm min}(x,y)\;.
\end{equation}
The overall potential on the entire $x$-axis is then given by a
periodically repeated realization of the BB (see
Fig. \ref{Fig1}).  Without loss of generality we set $l = 1$ in what
follows, meaning that we measure $L$ in units of $l$. We will also
skip insignificant numerical factors focusing only on the dependence
on the pertinent parameters, such as $T$, $L$ and $V_0$.

Before we proceed, it is important to emphasize that the dynamics in
Eq.(\ref{Langevin}) represents a combination of two paradigmatic
situations: random motion in a periodic potential and the Sinai
dynamics. Consequently, we expect that $x(t)$ will exhibit two
distinct temporal behaviors.  At sufficiently short times t, $t \ll
t_c$, where $t_c$ is a crossover time, the periodicity will
not matter and the evolution of $x(t)$ will proceed exactly in the
same fashion as in the original Sinai model, \eqref{sinai}.  At longer
times, $t \gg t_c$, the periodicity of the potential will ensure a
transition to a standard diffusive behavior, so that $x(t)$ will
converge to
\begin{equation}
\label{Diffusion}
  x(t) \sim \sqrt{2 D[V(x)]} \, B_t \,,
\end{equation}
where $B_t$ is a Brownian trajectory with diffusion coefficient $1$
and $D[V(x)]$ is a sample-dependent diffusion coefficient (see, {\em
  e.g.}, \cite{lif,ar,hang,dav}):
\begin{equation}
\label{diffusion}
  D[V(x)] = D_0/\left(\int^L_0 \dfrac{dx}{L} \int^L_0 \dfrac{dy}{L}
  \exp\left(\dfrac{V(x) - V(y)}{T}\right)\right)
\end{equation}
where $D_0 = T/\eta$ is the bare diffusion coefficient in absence of
disorder. Note that $D[V(x)] \leq D_0$ \cite{lif} so that $D[V(x)] $
is a random variable with support on $(0,D_0)$.

\begin{figure}[ht]
\begin{center}
\centerline{\includegraphics[width = .32 \textwidth,angle=270]{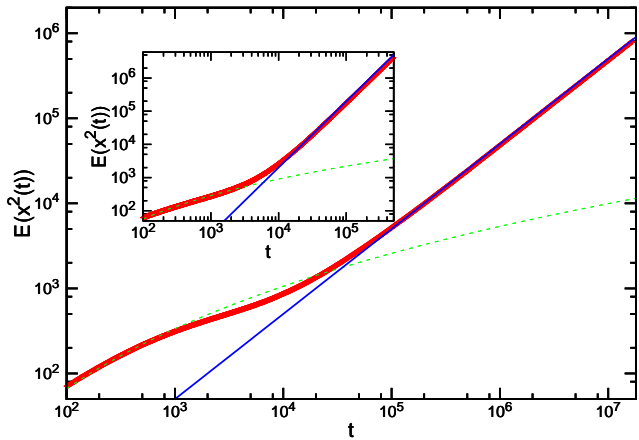}}
\caption{(color online) Main: $\mathbb{E}\left(\overline{x^2(t)}\right)$ in a
  periodic Sinai model (with a periodic BB), numerical data shown as points.
  Also shown by the dashed line is the fit $c_1 \ln^4(t)$ for the short time Sinai regime along with the solid line late time fit $c_2t$.  Inset: as in the main figure, 
  but for an unconstrained periodic Sinai potential. 
\label{Fig2}
}
\end{center}
\end{figure} 

In the main plot of Fig. \ref{Fig2} we show the temporal evolution of
$\mathbb{E}\left(\overline{x^2(t)}\right)$ in a periodic BB Sinai
model, with $L=64$. The numerical evidence for the existence of the
two temporal regimes described in \eqref{sinai} (at short times) and
in \eqref{Diffusion} (at large times) is clear.  We plot with points
the numerical data averaged over 500000 realizations of the random
quenched potential. The dashed line is $\ln^4(t)$
and agrees with the simulated data in the time region $(100,1000)$, while the continuous
 thin straight line is $t$ and fits perfectly the
asymptotically large time region (say from $t_{\mbox{min}}=10^5$). 

An intermediate very slow regime, where both the $\ln^4(t)$ and the
$t$ dependence fail to fit the data, also appears clearly. Such a
departure from the $\ln^4(t)$ law is not observed for a periodic
\textit{unconstrained} Sinai potential, that we show in the inset,
again for $L=64$ (here the transition is from a Sinai to a ballistic
regime, since for any finite $L$ the potential is biased yielding an constant, but random from sample to sample, force superimposed on a periodic potential).  
As a matter of fact, this is a surprising
feature since one may intuitively expect that in the case of a
BB potential the typical barrier which a particle has to
overcome should be less, due to stronger correlations, than that for
an unconstrained Brownian potential, so that for a BB the mean-square displacement
$\mathbb{E}\left(\overline{x^2(t)}\right)$ should grow faster with time. This
appears not to be the case and an apparent explanation is that for the
BB potential the structure of a typical barrier which a
particle has to bypass is different from the one for an unconstrained
Brownian motion.  This may be related to the  recent observation
\cite{gleb2} that the variance of a maximal positive displacement of a
BB on some sub-interval $(0,L_1)$ with $L_1 < L$, may be
greater than the variance of the maximal displacement on the entire
interval $(0,L)$.

The inset helps us noticing that the transition from the Sinai regime
at short times to the long time regime is not smeared in time but is
sharp, and allows to consistently define a well-defined value of a
transition time $t_c$, which we will discuss below. Accounting for the
intermediate, sub-diffusive regime that appears in the case of the
Sinai periodically repeated Brownian bridge, the same procedure allows
to define a transition time also in this case.  We may expect that for
$t \gg t_c$, the typical behavior of $x(t)$ will be diffusive, so that
the low-frequency ($f \ll 1/t_c$) behavior of the power spectrum
\eqref{spec1} will have the form of \eqref{spec2} with $\alpha = 2$
\begin{align}
\label{a}
  \dfrac{{\cal A}}{4 D_0} = \dfrac{1}{\int^L_0 \dfrac{dx}{L}
  \int^L_0 \dfrac{dy}{L} \exp\left(\dfrac{V(x) - V(y)}{T}\right) }\,.
\end{align}  
Taking into account that for a standard Brownian motion with the
diffusion coefficient $D$ the amplitude in \eqref{spec2} is
${\cal A} = 4 D$, we expect ${\cal A}$ to have support $(0,4 D_0)$. In
what follows we will focus on the statistical properties of
${\cal A}$.

We start by analyzing the typical behavior of ${\cal A}$ based on an
estimate for the typical value of $\cal A$ that we call
${\cal A}_{typ}$:
\begin{align}
  \dfrac{{\cal A}_{typ}}{4 D_0} \sim \exp\left(\mathbb{E}
    \left(\ln\left(\dfrac{{\cal A}}{4 D_0}\right)\right)\right).
\end{align} 
Furthermore, 
\begin{align}
  \mathbb{E}\left(\ln\left(\dfrac{{\cal A}}{4 D_0}\right)\right)
  = \mathbb{E}\left(\ln J_L^{+}\right)
  + \mathbb{E}\left(\ln J_L^{-}\right) + 2\ln(L)\;,
\end{align}
where $J_L^+$ and $J_L^-$ are stationary currents through a finite, of
length $L$ sample of a Sinai chain,
\begin{align}
\label{currents}
J_L^+ = \dfrac{1}{\int^{L}_0
  dx \exp\left(\dfrac{V(x)}{T}\right)} \,,\nonumber\\
J_L^- = \dfrac{1}{ \int^{L}_0
  dy \exp\left(-\dfrac{V(y)}{T}\right)} \,.
\end{align}
Note that since $ \mathbb{E}\left(V(x)\right) = 0$, moments of
arbitrary order obey $ \mathbb{E}\left(\left(J_L^{+}\right)^k\right)
\equiv \mathbb{E}\left(\left(J_L^{-}\right)^k\right)$ so that
\begin{align}
\mathbb{E}\left(\ln J_L^{+}\right) \equiv \mathbb{E}\left(\ln J_L^{-}\right)
\end{align}
and thus
\begin{align}
  \mathbb{E}\left(\ln\left(\dfrac{{\cal A}}{4 D_0}\right)\right)
  = 2 \mathbb{E}\left(\ln J_L^{+}\right) + 2\ln(L).
\end{align}
The statistical properties of the currents in finite Sinai chains have
been analyzed in \cite{osh1,osh2,mon,alb} for the case where $V(x)$ is
an unconstrained Brownian or an unconstrained fractional Brownian
motion. It was shown (see, e.g. \cite{alb} for more details) that for
sufficiently large values of $L$, the behavior of $J_L^+$ is dominated
by the maximum of $V(x)$,
$V_{max} \equiv {\rm max}_{0 \leq x \leq L}V(x)$.  Moreover, for any
given realization of disorder $J_L^+$ can be bounded from below and from
above by $A_1 \exp(- V_{max}/T)$ and $A_2 \exp(- V_{max}/T)$, where
$A_1 \leq A_2$ are $L$-independent constants.  Consequently, the
$L$-dependence (up to an insignificant numerical factor) is captured
by the estimate $J_L^+ \sim \exp(- V_{max}/T)$.

In principle, this
argument can be readily generalized 
for the case at hand, when $V(x)$
is a BB, and we have merely to use the distribution
$P_{BB}(V_{max})$ of a maximal positive displacement of a BB on an interval $(0,L)$, instead of the analogous distribution
for an unconstrained Brownian motion used in \cite{alb}. This
distribution $P_{BB}(V_{max})$ is well-known from the classical papers
\cite{kol,sm,fel}, and is given by
\begin{align}
\label{bb}
P_{BB}(V_{max}) = \dfrac{2 V_{max}}{D_V L}
\exp\left(- \dfrac{V_{max}^2}{D_V L}\right) \,,
\end{align}
where $D_V = V_0^2/(2 l)$.
Using \eqref{bb}, we find that, dropping numerical constants, 
\begin{align}
  \mathbb{E}\left(\ln\left(\dfrac{{\cal A}}{4 D_0}\right)\right)
  \sim - \dfrac{V_0}{T} L^{1/2} \,,
\end{align}
so that, for arbitrary values of $k$, 
\begin{align}
\label{typ}
  \left(\dfrac{{\cal A}_{typ}}{4 D_0}\right)^k \sim \exp\left(- k
  \dfrac{V_0}{T} L^{1/2} \right) \,.
\end{align}
Therefore, we expect that, for most realizations of the random
potential $V(x)$, the amplitude ${\cal A}$ of the power spectrum will
decrease, as a stretched-exponential function $\exp(- L^{1/2})$ of the
periodicity $L$, and will exhibit an Arrhenius dependence on the
temperature $T$.

Next we consider the behavior of the moments $\mathbb{E}({\cal A}^k)$
of the amplitude with arbitrary (positive or negative) values of
$k$. When $V(x)$ is an \textit{unconstrained} Brownian motion, a
general analysis of the functional in \eqref{diffusion} or \eqref{a}
has been presented in \cite{dean}. The disorder-average value
(first moment) of this very functional, which also describes the
ground-state energy in a one-dimensional localization problem,
was determined in \cite{burl}. It was shown in \cite{dean,burl} that
the functional of the random potential in \eqref{diffusion} or
\eqref{a} can be bounded from below and from above by $B_1 \exp(-R/T)$
and $B_2 \exp(- R/T)$, where $B_1 \leq B_2$ weakly depend on $L$ and
\begin{align}
R \equiv {\rm max}_{0 \leq x \leq L} V(x) - {\rm min}_{0 \leq x \leq L} V(x) 
\end{align}
is the range, or span, of the random potential $V(x)$. Physically $R$
corresponds to the largest energy barrier that will be encountered by
the tracer. Expecting that $\mathbb{E}({\cal A}^k)$ will show a stronger than a power-law dependence on $L$ (and we will show in what follows that it is the case) we may drop the constants $B_1$ and $B_2$ and write an estimate
\begin{equation}
\label{A}
{\cal A}^k \sim \exp(- k R/T) \,,
\end{equation}
which should capture the $L$, $k$ and $T$ dependence of the moments up to 
insignificant
pre-exponential factors.

To extend this analysis over the case of a BB potential and
 in order to calculate the moments of ${\cal
  A}$ for the case under study, we need to know the distribution of the range of a
BB. This distribution was first derived in
\cite{feller}, in which $R$ was referred to as {\em an adjusted range}
of Brownian motion, and it is given in series form as
\begin{align}
\label{d1} 
  &P_{BB}(R) = R \dfrac{d^2 f(R)}{dR^2}
+ \sum_{n=2}^{\infty} \Bigg[2 n(n - 1)
  \Bigg(\dfrac{d  f((n-1) R)}{dR} - \nonumber\\
  &- \dfrac{d f(n R)}{dR} \Bigg) 
  + (n-1)^2 R  \dfrac{d^2 f((n-1) R)}{dR^2}
  + n^2 R \dfrac{d^2 f(n R)}{dR^2} \Bigg] \,,
\end{align}
where, in our notation, $f(R) = \exp(-R^2/D_V L)$.  For our purposes a
slightly different form of $P_{BB}(R)$ will also turn out to be useful. To
this end, we exploit here the observation made in \cite{ken} that the
range of Brownian Bridge and the maximum of Brownian excursion - a
Brownian Bridge constrained to stay positive - have the same
distributions. The distribution of the maximum of a Brownian excursion
has been extensively discussed in the literature and several forms of
it have been derived (see for example \cite{yor}).  Choosing a
suitable one, we have, in our notation,
\begin{align}
\label{d2}
&P_{BB}(R) = \sqrt{2} \pi^{5/2} (2 D_V L)^{3/2} \times  \nonumber\\ &\dfrac{d}{d R}
  \left(\dfrac{1}{R^3} \sum_{n=1}^{\infty} n^2
\exp\left(- \dfrac{\pi^2 n^2}{R^2} D_V L\right)\right) \,.
\end{align} 
The two expressions \eqref{d1} and \eqref{d2} coincide.

Now we have all necessary ingredient to calculate the moments of
${\cal A}$. Consider first the moments of negative (not necessarily
integer) order.  Using the form of $P_{BB}(R)$ in \eqref{d1}, and
keeping only the leading exponential dependence on $R$, we average the estimate in \eqref{A}
 to obtain 
\begin{align}
  \mathbb{E}\left(\left(\dfrac{4 D_0}{{\cal A}}\right)^k\right)
  \sim \int^{\infty}_0 dR \exp\left(\dfrac{k \, R}{T} -
  \dfrac{R^2}{D_V L}\right) 
\end{align}
Evaluating this integral via steepest descent, we find that the
maximum of the exponential is attained at $R \sim R^* = k D_V L/2T$,
and thus
\begin{align}
\label{mommin}
\mathbb{E}\left(\left(\dfrac{4 D_0}{{\cal A}}\right)^k\right)
\sim \exp\left(\dfrac{k^2 \, V_0^2}{8T^2} L\right) \,.
\end{align}
Therefore, the negative moments grow faster than exponentially with
$k$ and $V_0$, exhibit a \textit{super}-Arrhenius dependence on the temperature
and grow exponentially with the periodicity $L$. 

The negative moments may also be computed directly by taking the
average over the replicated 2$k$-fold integral to obtain
\begin{align}
&\mathbb{E}\left(\left(\dfrac{4 D_0}{{\cal A}}\right)^k\right) = 
 \int_0^1\ldots \int^1_0 \prod_{a=1}^k du_adw_a \times \nonumber\\
&  \exp\Bigg( -\frac{D_VL}{2T^2}
    \Bigg(\sum_{a,b} |u_a-u_b| +   |w_a -w_b| \nonumber\\
    &-2|u_a-w_b|
    + 2\left(\sum_a u_a -\sum_a w_a\right)^2\Bigg)\Bigg),
    \label{part}
\end{align}
where we have rewritten the integration variables using $x_a = L u_a$
and $y_a = Lw_a$ to obtain the above. The right hand side
of \eqref{part} has the form of a partition function for $k+k$
interacting particles of two types $u$ and $w$ at inverse {\em
  temperature} $\beta = D_VL/2T^2$. In the limit of large $L$
the partition function is dominated by the ground state energy. The
Hamiltonian is explicitly given by
\begin{equation}
  H = \sum_{a,b} |u_a-u_b| + |w_a -w_b| -2|u_a-w_b|
  + 2(\sum_a u_a -\sum_a w_a)^2\;.
\end{equation}
The particles of type $u$ and $w$ attract particles of the same type with a
linear attractive potential, and they repel particles of the other
type, again with a linear potential. However there is an additional
interaction which harmonically binds the center of masses of the two
particle types. Due to the attraction between the same particle type
we expect that particles of the same type will condense at low temperature
about the same point and hence we write
$u_a = U$ and $w_a = W$ for all $a$. This gives the effective reduced
low temperature Hamiltonian
\begin{equation}
H_0 = 2k^2 (\Delta^2-\Delta ) = 2k^2 (\Delta-\frac{1}{2})^2 - \frac{k^2}{2}\;,
\end{equation}
where $\Delta = |U-W|$. The value $\Delta = 1/2$ minimizes the energy
leading to
\begin{equation}
  \mathbb{E}\left(\left(\dfrac{4 D_0}{{\cal A}}\right)^k\right)
  \sim \exp(\frac{D_Vk^2L}{4T^2})
  = \exp(\frac{k^2V_0^2L}{8T^2})\;,
\end{equation}
in complete agreement with \eqref{mommin}.

For positive moments of the amplitude, we use the form of $P_{BB}(R)$
in \eqref{d2}.  Keeping only the leading term in $L$, we find that the
leading behavior  of
${\cal  A}^k$ in \eqref{A} is given by
\begin{align}
  \mathbb{E}\left(\left(\dfrac{{\cal A}}{4 D_0}\right)^k\right)
  \sim \int^{\infty}_0 dR \exp\left(-\dfrac{k \, R}{T} -
  \dfrac{\pi^2 D_V L}{R^2}\right) \,.
\end{align} 
Again, we use the steepest descent approach to observe that the
dominant contribution to the integral comes from a narrow region
around $R^* = (2\pi^2T D_V L/k)^{1/3}$ so that the overall behavior of
the positive moments of the amplitude of (not necessarily integer)
order $k$ is given by
\begin{align}
\label{momplus}
\mathbb{E}\left(\left(\dfrac{{\cal A}}{4 D_0}\right)^k\right)
\sim \exp\left(- \frac{3 \, \pi^{\frac{2}{3}}}{2}
\left(\dfrac{k \, V_0}{T}\right)^{2/3} \, L^{1/3}\right) \,,
\end{align}
Therefore, the positive moments of the amplitude exhibit a
stretched exponential dependence on the order of the moment $k$ and on
the characteristic scale of the potential $V_0$, a \textit{sub}-Arrhenius 
dependence on the temperature,
and also decay with
the periodicity $L$ as a stretched exponential with the exponent
$z=1/3$, that is to say, {\em slower} than predicted by the estimate
based on the typical realizations of disorder, \eqref{typ}. Note,
however, that the result in \eqref{momplus} pertains to the asymptotic
limit when $L \to \infty$. For small values of $L$ we expect that
positive moments will exhibit the typical behavior given by \eqref{typ}.

\begin{figure}[ht]
\begin{center}
\centerline{\includegraphics[width = .32 \textwidth,angle=270]{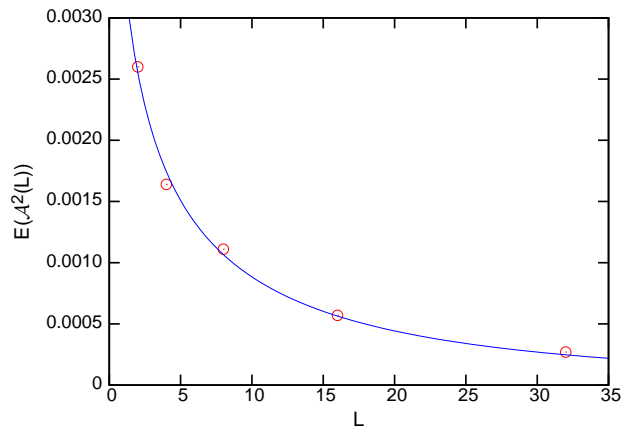}}
\caption{(color online) The second moment of the fitted
amplitudes of the power spectrum in a BB potential. Numerical results are shown
as circles along with the fit of the data by $a\exp(- L^b)$, shown as a solid line, yielding the fitted
value $b=0.37$}
\label{Fig3}

\end{center}
\end{figure}

In Fig.~\ref{Fig3} we show with symbols our estimates for ${\cal A}^2$
obtained from numerical simulations for different values of $L$. In
this case we are not able, in the limits of our numerical precision,
to distinguish a small $L$ regime. The continuous line is our best fit
to the form $a\exp(-L^b)$, where we obtain the value $b=0.37\pm
0.02$. The precision of the numerical data does not support a fit with
more parameters (that means that we cannot include subleading
corrections). The value $b=0.37$ that we find for our estimated exponent
(in the sense it is estimated by numerical data in a finite region of
values of $L$) is close to the expected asymptotic value of $1/3$ for large
$L$, but probably feels the contamination from the low $L$ regime.

Before we proceed to the analysis of the distribution of the amplitude
${\cal A}$, two remarks are in order. We first note that there exists another physical system, completely
unrelated to the one under study, which exhibits essentially the same
behavior. It concerns survival of diffusing particles, with diffusion
coefficient $D_V$, in presence of perfect traps, independently and 
uniformly distributed on a one-dimensional line. Identifying $L$ as {\em time} and
$1/T$ as the {\em density} of traps, we see that in one-dimensional
systems the behavior of the moments of the probability $S_L$ that a
particle {\em survives} up to time $L$ is identical to the behavior of
the moments of ${\cal A}$ (see, e.g., \cite{yuste} and references
therein). At sufficiently short times $L$, $S_L$ follows the
stretched-exponential form in \eqref{typ}, which is tantamount to the
so-called Smoluchowski regime, while for $L \to \infty$, the moments
of $S_L$ obey the form in \eqref{momplus} as they are supported by the
{\em optimal fluctuation} $R^* = (T V_0^2 L/k)^{1/3}$ of a random
cavity devoid of traps. This ultimate, late time, regime has the
celebrated fluctuation-induced tails \cite{bal,don}, which are also
intimately related to the so-called Lifshitz singularity in the
low-energy spectrum of an electron in a one-dimensional disordered
array of scatterers \cite{lifsh}. Below we will show that an analogous
essential singularity shows up in the distribution $P({\cal A})$.

Secondly, we are now in position to estimate the crossover time $t_c$,
and hence, to determine the upper bound on the frequency for which the
spectrum \eqref{spec2} is characterized by an exponent
$\alpha = 2$. Recalling that our
numerical results show a sharp crossover from the Sinai regime
\eqref{sinai} to the diffusive behavior in \eqref{Diffusion}, we may
estimate $t_c$ by simply equating the mean squared displacement in the
Sinai \eqref{sinai} and diffusive regimes \eqref{Diffusion}, {\em
  i.e.}
\begin{align}
  \ln^4(t_c) \sim \mathbb{E}\left(D[V(x)]\right) \, t_c \,,
\end{align}
which gives
\begin{align}
t_c \sim \dfrac{1}{\mathbb{E}\left(D[V(x)] \right)} \,.
\end{align}
Now noticing that $D[V(x)] \sim {\cal A}$, we can expect that $t_c$
will display a different dependence on the periodicity $L$ (and the
other system parameters) for small and large values of $L$. For
sufficiently small $L$ (but still large enough so that the behavior in
\eqref{sinai} has enough space to emerge), the typical trajectories of
disorder, such that $|V(x)| \sim \sqrt{x}$, will dominate and
\begin{align}
\label{tc1}
  t_c \sim \exp\left(\dfrac{V_0}{T}  L^{1/2}\right) \,,
\end{align}
which simply tells us that, for sufficiently small $L$, the crossover
time $t_c$ to diffusive regime is a time needed for $x(t)$ to travel
over a distance $L$ encountering a typical barrier $V_0 L^{1/2}$ which
$x(t)$ overcomes due to thermal activation.  Note the Arrhenius
dependence of $t_c$ on the temperature $T$.

For larger values of $L$ the behavior of the average 
amplitude ${\cal
  A}$, given by Eq. \eqref{momplus}, becomes
supported by atypical realizations of disorder with the {\em
  optimal fluctuation} trajectories of $|V(x)| \sim x^{1/3}$.  For
such $L$, we have, by virtue of \eqref{momplus},
\begin{align}
  \label{tc2}
  t_c \sim \exp\left(c\left(\dfrac{V_0}{T}\right)^{2/3} L^{1/3} \right) \,,
\end{align}
where $c$ is a numerical constant; this means that, for larger
periodicities, $t_c$ exhibits a slower growth with $L$.  Note that in
this case $t_c$ has a rather unusual sub-Arrhenius dependence on the
temperature.  

In order to discuss this point and to use our numerical data to better
understand it, we start by defining a time of exit from the Sinai
asymptotic regime. The Sinai regime holds in the first part of the
dynamical evolution. We define an exit time from it as the time 
$t_c^{(1)}$ as the minimal time such that
\begin{equation}
\mathbb{E}\left( \overline{|x(t)|} \right) - \mathbb{E} \left( \overline{|x(t)|} \right)_{Sinai}
  >  3 \sigma_{Sinai}(t)\;,
\end{equation}
where by the $Sinai$ label we denote an average over the motion in
an infinite, unconstrained Sinai potential. In this way we are
observing the time where the departure of the motion in the
periodic Brownian Bridge potential is  substantially different from
the one in a Sinai infinite potential ($\sigma_{Sinai}$ is the
standard deviation over our numerical estimate for the infinite Sinai
motion). On our time scales and sample size this procedure is accurate
enough to give a sensible  estimate of $t_c^{(1)}$. We assume now that
\begin{equation}
\ln(t_c^{(1)})   \sim a^{(1)} +  L^{b^{(1)}}\;.
\label{eq:tc1vsL}
\end{equation}
Since our numerical data are not accurate enough to allow  us to
disentangle precisely the subleading corrections to this behavior, we
analyze our data by defining a size dependent exponent $b^{(1)}(L,2L)$,
computed by using Eq. (\ref{eq:tc1vsL}) for size $L$ and size $2L$.
The numerical values computed
for $t_c^{(1)}(L)$ and the one for  $t_c^{(1)}(2L)$ are used to disentangle the
value of $b^{(1)}(L,2L)$ as estimated from these two values of the lattice size.
The limit for large $L$ of  $b^{(1)}(L,2L)$  is  $b^{(1)}$.

\begin{figure}[ht]
\begin{center}
\centerline{\includegraphics[width = .32 \textwidth,angle=270]{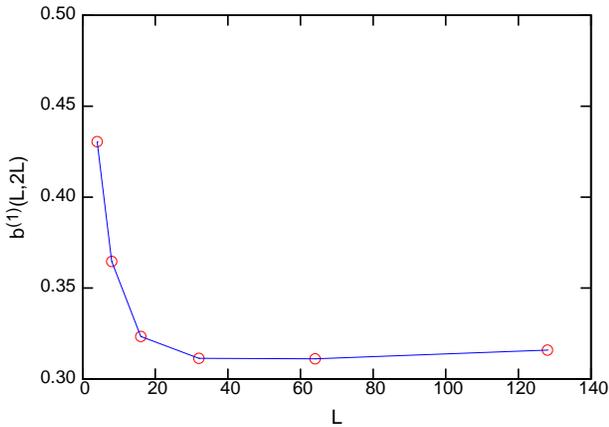}}
\caption{(color online) The exponent
$b^{(1)}(L,2L)$ 
in Eq.(\ref{eq:tc1vsL})
as a function of $L$. 
\label{Fig4}
}
\end{center}
\end{figure}

We plot
this estimated exponent as a function of $L$ in
Fig. \ref{Fig4}. In this case the crossover we have derived
analytically clearly emerges from the numerical data, that give an
estimated exponent close to $1/2$ for small $L$ values and close to
$1/3$ for larger values of the size $L$. 

\begin{figure}[ht]
\begin{center}
\centerline{\includegraphics[width = .5 \textwidth,angle=0]{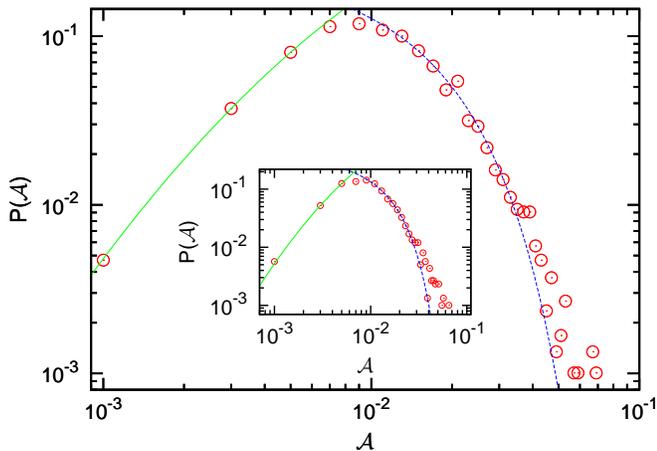}}
\caption{(color online) Distribution $P({\cal A})$ of the amplitudes ${\cal A}$ for a BB potential, 
plotted with circles (numerical results). In the main figure $L=64$ while the inset shows the results for $L=32$. The log-normal fit corresponding to Eq. (\ref{ln}) is shown for small ${\cal A}$ as a solid green curve while the prediction of Eq. (\ref{d4}) for the right tail is shown 
for large values of ${\cal A}$ by the dashed blue line.
\label{Fig5}
}
\end{center}
\end{figure}

We finally turn to the analysis of the distribution $P({\cal A})$ of
the amplitude of the low-frequency power spectrum (see Fig. \ref{Fig5}).
Examining first the negative moments of ${\cal A}$, we observe that
they are growing functions of $L$ and $k$, which hints that such a
behavior of ${\cal A}$ is derived from the left-tail of the
distribution $P({\cal A})$, {\em i.e.}, when ${\cal A}$ is close to
$0$. Furthermore, the quadratic dependence of the moments on the order
of the moment $k$ in the exponential is a fingerprint of the
log-normal distribution, which suggest that the left-tail of
$P({\cal A})$ has the form:
\begin{align}
\label{ln}
P({\cal A}) \sim \dfrac{1}{{\cal A}} \exp\left(- \dfrac{2T^2 \ln^2
  \left({\cal  A}\right)}{V_0^2 L}\right) \,.
\end{align}
Note that this distribution is uni-modal, with the most probable value
of ${\cal A}_{mp} \sim \exp(- V_0^2 L/4T^2)$, which is, for
sufficiently large $L$, much smaller and closer to $0$ than the {\em
  typical} value in \eqref{typ}.
Further on, positive moments in \eqref{momplus} are,
for large $L$, much larger than those expected
from the typical realizations of disorder, \eqref{typ}.  This means,
in turn, that the behavior in \eqref{momplus} stems apparently from
the right-tail of the distribution $P(\cal A)$ when ${\cal A}$ is
close to the right edge of the support, i.e., ${\cal A} \approx A_r =
4 D_0$. Let us formally write
\begin{align}
\label{int}
  \int^{4 D_0}_0 {\cal A}^k d{\cal A} \, P({\cal A})
  \sim A_r^k \, \exp\left(- \left(\dfrac{k V_0
    \sqrt{L}}{T}\right)^{2/3}\right) \,,
\end{align}
where for simplicity of notation any numerical constant in the
exponential of the right hand side is included in $V_0$.  We
assume that the major contribution to the integral on the left hand
side of \eqref{int} comes from a narrow region close to the right edge
of the support. Changing the integration variable as
\begin{align}
  z = \dfrac{T}{V_0 \sqrt{L}} \ln\left(\dfrac{A_r}{{\cal A}}\right) \,,
\end{align} 
we cast \eqref{int} into the form
\begin{align}
&  \dfrac{4 D_0 V_0 \sqrt{L}}{T} \int^{\infty}_0 dz \,
  \exp\left(- \left(\dfrac{k V_0 \sqrt{L}}{T}\right) z\right) \, \times \nonumber\\
&  \exp\left(- \dfrac{V_0 \sqrt{L} z}{T}\right) \,
  P(z) \sim \exp\left(- \left(\dfrac{k V_0
      \sqrt{L}}{T}\right)^{2/3}\right) \,. 
\end{align}
Using then the formal definition of the Laplace transform of
one-sided stable L\'evy distribution ${\cal L}_{\nu}(z)$ with index
$\nu$ (see, e.g., \cite{greg})
\begin{align}
  \int^{\infty}_0 \exp\left(- p z\right) {\cal L}_{\nu}(z)
  \equiv \exp\left(- p^{\nu}\right)
\end{align}
we immediately infer that 
\begin{align}
\label{levy}
P({\cal A}) \sim \dfrac{T}{V_0 \sqrt{L} {\cal A}} {\cal L}_{2/3}
\left(\dfrac{T}{V_0 \sqrt{L}}
\ln\left(\dfrac{A_r}{{\cal A}}\right)\right) \,.
\end{align}
Note that the result in \eqref{levy} is expected to hold only in the
vicinity of the right edge of the support, and we consider its
asymptotic form in this domain. For ${\cal A} \approx A_r$, the
argument $z$ in the one-sided L\'evy distribution ${\cal L}_{2/3}(z)$
is close to zero, so that its asymptotic behavior is given by
\begin{align}
{\cal L}_{2/3}(z) \sim z^{-2} \exp\left(- \dfrac{b}{z^2}\right) \,,
\end{align}
where $b$ is a computable constant. For ${\cal A} \approx A_r$, we have that
\begin{align}
z \approx \dfrac{T}{V_0 \sqrt{L}} \left(1 - \dfrac{{\cal A}}{A_r}\right) \,,
\end{align}
so that eventually we find the following asymptotic representation of
the distribution $P({\cal A})$ close to the right edge of the support
\begin{align}
\label{d4}
P({\cal A}) \sim \dfrac{4 D_0 V_0 
\sqrt{L}}{\left(4 D_0 - {\cal A}\right)^2} 
\exp\left(- \left(\dfrac{4 D_0 V_0 \sqrt{b L}}{T 
\left(4 D_0 - {\cal A}\right)}\right)^2\right) \,.
\end{align}
Note that the distribution in \eqref{d4} exhibits an essential
singularity in the vicinity of $A_r$, which is related to the Lifshitz
singularity. In Fig.~\ref{Fig5} we plot the empirical probability
distribution obtained in numerical simulations, together with the best
fits to the asymptotic forms (\ref{ln}) and (\ref{d4}): the agreement
is remarkable.

EM and GO wish to thank the Kavli Institute for Theoretical Physics of
the University of California at Santa Barbara for warm hospitality
during their participation in March 2014 at the program {\em Active
  Matter: Cytoskeleton, Cells, Tissues and Flocks}, where this work
has been initiated.  DSD and GO acknowledge a partial support from the
Office of Naval Research Global Grant N62909-15-1-C076 and also thank
the Institute for Mathematical Sciences of the National University of
Singapore for warm hospitality and a financial support.  EM has used
towards development of this project funding from the European Research
Council (ERC) under the European Union’s Horizon 2020 research and
innovation programme (grant agreement No [694925]).

\end{document}